\documentclass[journals]{IEEEtran}
\usepackage{xspace,amsmath,amssymb,amsfonts,epsfig,subfigure,syntonly}
\usepackage{cite,bm,color,url,textcomp,empheq,boxedminipage}
\usepackage{algorithmicx,algorithm}
\usepackage{epstopdf,makecell}
\usepackage{empheq}
\usepackage{pifont}
\usepackage{subeqnarray}
\usepackage{cases}
\usepackage{graphicx,graphics}  % Written by David Carlisle and Sebastian Rahtz
\usepackage{multirow,multicol}
\usepackage{psfrag}    % Written by Craig Barratt, Michael C. Grant,
\usepackage{stfloats}
\usepackage{url}
\usepackage[noend]{algpseudocode}

\newtheorem{proposition}{\underline{Proposition}}[section]

\newcommand{\mv}[1]{\mbox{\boldmath{$ #1 $}}}

\psfull

\begin{document}
\title{Radio-Map-Based Robust Positioning Optimization for UAV-Enabled Wireless Power Transfer}

\author{{Xiaopeng Mo, Yuwei Huang, and Jie Xu\vspace{-20pt}}
\thanks{\scriptsize{X. Mo and J. Xu are with the School of Information Engineering, Guangdong University of Technology, Guangzhou, 510006, China (e-mail: xiaopengmo@mail2.gdut.edu.cn, jiexu@gdut.edu.cn). J. Xu is the corresponding author.}}
\thanks{\scriptsize{Y. Huang is with the School of Information Science and Technology,
University of Science and Technology of China, Hefei, 230027, China (e-mail:
hyw1023@mail.ustc.edu.cn)}}
}

\maketitle
\begin{abstract}
 This letter studies an unmanned aerial vehicle (UAV)-enabled wireless power transfer (WPT) system, in which a UAV-mounted energy transmitter (ET) optimizes its positioning locations over time to efficiently charge a set of energy receivers (ERs) distributed on the ground. Different from conventional designs based on deterministic (e.g., line-of-sight (LoS)) or stochastic (e.g.,  probabilistic LoS) channel models, we consider a new radio-map-based design approach, in which the UAV exploits the information of channel propagation environments for efficient positioning optimization. By practically assuming that the UAV only partially knows the ERs' locations, our objective is to maximize the minimum energy transferred to all ERs over a particular charging duration that is sufficiently long.
 By applying the robust optimization and Lagrange duality method, we obtain an efficient solution to the minimum energy maximization problem, which has an interesting multi-location-positioning structure. Numerical results show that our proposed radio-map-based robust design significantly improves the WPT performance, as compared to conventional designs based on LoS and probabilistic LoS channel models.
\end{abstract}
\vspace{-15pt}\section{Introduction}
Motivated by recent advancements in both unmanned aerial vehicles (UAVs) \cite{zeng2016wireless} and wireless power transfer (WPT) \cite{zeng2017communications}, UAV-enabled WPT has emerged as a new technique to provide convenient and cost-effective energy supply for massive low-power devices in future Internet-of-things (IoT) networks \cite{xu2018uav}. In this technique, UAVs are dispatched as aerial energy transmitters (ETs) to wirelessly charge energy receivers (such as sensors and IoT devices) distributed on the ground.  Recently, UAV-enabled WPT has also been extended to other applications such as UAV-enabled wireless powered communication networks \cite{xie2018throughput,S.cho2018,M.Jiang2019} and wireless-powered mobile-edge computing\cite{F.Zhou.2018}.

By exploiting UAVs' controllable mobility, positioning and trajectory designs have been recognized as an important solution for the UAV to increase the transferred energy amounts towards multiple ground energy receivers (ERs).
In the UAV positioning/trajectory design literature, prior works normally assumed line-of-sight (LoS) (see, e.g.,\cite{xu2018uav,xie2018throughput}) or probabilistic LoS channel models (see, e.g., \cite{al2014optimal,W.Saad2016}) for air-to-ground wireless links, and also assumed that the UAV perfectly knows the exact locations of ground nodes (ERs of our interest). These assumptions, however, may not be true in practice. First, the deterministic LoS channel model is only applicable in open areas with no obstacles around ERs, while the probabilistic LoS channel model only captures the stochastic property of wireless environments over a large area. For practical scenarios with obstacles like trees or buildings around, these models cannot characterize the exact channel propagation environments. Next, due to the limited accuracy of practical localization methods (e.g., global positioning systems (GPS)), the UAV may only partially know the ERs' location information with certain errors. Thus, how to optimize the UAV's positioning or trajectory design under practical channel setups with imperfect ERs' location information is a challenging problem that is not well addressed yet.

In particular, this letter studies the positioning optimization in a UAV-enabled WPT system over a sufficiently long charging duration{\footnote{In general, for a practical WPT area of tens or hundreds of square meters, the charging duration (e.g., several or tens of minutes) is significantly longer than the duration for the UAV to fly around this area (e.g., several seconds). Therefore, it is practically relevant to assume the charging duration to be sufficiently long, such that the flying duration is negligible. }}, in which one UAV optimizes its positioning locations over time to efficiently charge multiple ERs on the ground. It is assumed that the UAV partially knows the ERs' locations, subject to norm-bounded errors. Furthermore, different from prior works considering LoS or probabilistic LoS channels, we consider a generic channel model, and suppose that the UAV is aware of the exact channel propagation information by using the radio map technique \cite{esrafilian2018learning,bi2019engineering,J.Chen2017}. Under this setup, we aim to maximize the minimum of the energy transferred to all ERs by optimizing the UAV's  positioning locations over time. Although this problem is non-convex and difficult to solve in general, we present an efficient algorithm to find a high-quality multi-location-positioning solution via the robust optimization and Lagrange duality method. Numerical results show that by efficiently exploiting the channel propagation information, our proposed radio-map-based robust design achieves significantly higher WPT performance, as compared to conventional designs based on LoS and probabilistic LoS channel models.
\vspace{-10pt}\section{System Model}
In this letter, we consider a UAV-enabled multiuser WPT system, where a UAV is dispatched as an aerial ET to wirelessly charge a set $\mathcal{K} \triangleq \{1,...,K\}$ of ground ERs over a given duration $T$ that is sufficiently long. Let $\mv{w}_k=(x_k,y_k)$ denote the horizontal location of ER $k\in \mathcal{K}$.
%Suppose that each ER $k\in\mathcal{K}$ is fixed at the three-dimensional coordinate $(x_k,y_k,0)$, with $\mv{\omega}_k=(x_k,y_k)$ denoting its corresponding horizontal location of ER $k\in\mathcal{K}$.
It is assumed that the UAV only knows the approximated location of each ER $k$, denoted by $\mv{\bar{w}}_k=(\bar{x}_k,\bar{y}_k)$, with the maximum location error being $\epsilon>0$. We thus have $\mv{w}_k\in\mathcal{A}_k=\{\mv{w}_k| \parallel\mv{w}_k-\mv{\bar{w}}_k\parallel\leq\epsilon\},\forall k\in \mathcal{K}$, where $\|\cdot\|$ denotes the Euclidean norm of a vector.
 The UAV is assumed to fly at a constant altitude $H>0$ in meter (m) with the horizontal location $\mv{q} (t)= (x(t),y(t))$ at time $t$, which is time-varying in general due to the UAV's mobility. As the charging duration $T$ is sufficiently long, we omit the flying time from one positioning location to another, and only consider the optimization of positioning or hovering locations and their corresponding durations.
Accordingly, the distance between the UAV and ER $k\in\mathcal{K}$ at time $t$
 is denoted as $d(\mv{q}(t),\mv{w}_k)=\sqrt{H^2+\|\mv{q}(t)-\mv{w}_k\|^2}$ .

We consider a generic path loss model, in which the channel power gain from the UAV to ER $k\in\mathcal{K}$ at time $t$ is given by
\begin{align}
h_{k}(\mv{q}(t),\mv{w}_k)
&=\frac{\beta(\mv{q}(t),\mv{w}_k)}{d(\mv{q}(t),\mv{w}_k)^{\alpha(\bm{q},\bm{w}_k)}}\nonumber\\
&=\frac{\beta(\mv{q}(t),\mv{w}_k)}{(H^2+\|\mv{q}(t)-\mv{w}_k\|^2)^{\alpha(\bm{q}(t),\bm{w}_k)/2}},
\end{align}
where $\beta(\mv{q}(t),\mv{w}_k)$ denotes the channel power gain at a reference distance of $d_0$ = 1 m, and $\alpha(\mv{q}(t),\mv{w}_k)$ denotes the path loss exponent.
% For example, the UAV can learn the radio map (i.e., the path loss exponent and the reference channel power gain at different locations) in advance according to various methods, such as machine learning technique.
%%\renewcommand\theremark{\arabic{section}.\arabic{remark}}
 %\begin{remark}
Notice that under our generic model, the parameters $\beta(\mv{q}(t),\mv{w}_k)$ and $\alpha(\mv{q}(t),\mv{w}_k)$ are dependent on the UAV's location $\mv{q}(t)$ and ER $k$'s location $\mv{w}_k$, due to different propagation conditions over this area. It is assumed that for any given $\mv{w}_k$, when the distance $d(\mv{q}(t),\mv{w}_k)$ increases, the parameter $\beta(\mv{q}(t),\mv{w}_k)$ decreases monotonically and $\alpha(\mv{q}(t),\mv{w}_k)$ increases monotonically, thus leading to larger path loss. This is practically relevant, due to the fact that if $d(\mv{q}(t),\mv{w}_k)$ increases, then the elevation angle decreases, and accordingly, there will potentially exist more obstacles blocking the communication link \cite{esrafilian2018learning,J.Chen2017}. Notice that our considered channel model in (1) captures the segmented path loss model in \cite{J.Chen2017} , as well as the LoS \cite{xu2018uav,xie2018throughput,S.cho2018,M.Jiang2019,F.Zhou.2018} and probabilistic LoS path loss models \cite{al2014optimal,W.Saad2016,esrafilian2018learning} as special cases. To exploit the channel propagation information, we consider that the UAV can adopt the radio map technique \cite{bi2019engineering}  to efficiently acquire the detailed geographical channel information (i.e., the path loss exponent $\alpha(\mv{q}(t),\mv{w}_k)$ and the reference channel power gain $\beta(\mv{q}(t),\mv{w}_k)$ under different locations). In practice, such information can be obtained by the UAV or other cooperating nodes {\it  a priori} via spectrum sensing together with machine learning techniques \cite{esrafilian2018learning}.

Under this setup, the transferred radio frequency (RF) power towards each ER $k\in\mathcal{K}$ at time $t$ is  given by
 \begin{eqnarray}
Q_{k}(\mv{q}(t),\mv{w}_k)=h_{k}(\mv{q}(t),\mv{w}_k)P
=\frac{\beta(\mv{q}(t),\mv{w}_k)P}{d(\mv{q}(t),\mv{w}_k)^{\alpha(\bm{q}(t),\bm{w}_k)}},
\end{eqnarray}
where $P > 0$ denotes the fixed transmit power at the UAV.
%However, in fact, it is difficult for the UAV to obtain the exact locations of ground ERs, due to the limitation of measurement, such as the location measurement accuracy of GPS.  By contrast, the UAV only can obtain the approximated location of each ER k, denoted by $\mv{\bar{w}}_k=(\bar{x}_k,\bar{y}_k)$. Denote $\mv{\delta}_k=(\Delta x_k,\Delta y_k)$ as the error between the ER $k$' exact location $\mv{w}_{k}$ and approximated location $\mv{\bar{w}}_{k}$, we have $|\mv{\delta}_{k}|\leq \epsilon$, where $\epsilon$ denotes the maximum location error.
% As a result, we have $\mv{w}_k=\mv{\bar{w}}_k+\mv{\delta}_k,\forall k\in \mathcal{K}$.
% By substituting new $\mv{w}_k$ into (3), the received power of each ER $k\in\mathcal{K}$ can be re-expressed as
%\begin{align}
%Q_k(\mv{q},\mv{\bar{w}}_{k}+\mv{\delta}_k)
%&=\frac{\eta\beta(\mv{q},\mv{\bar{w}}_k+\mv{\delta}_k)P}
%{d_k(\mv{q},\mv{\bar{w}}_{k}+\mv{\delta}_k)^{\alpha(\mv{q},\mv{\bar{w}}_k+\mv{\delta}_k)}}\nonumber\\
%&=\frac{\eta\beta_(\mv{q},\mv{\bar{w}}_k+\mv{\delta}_k)P}{(H^2+\|\mv{q}-(\mv{\bar{w}}_k+\mv{\delta}_k)\|^2)^{\alpha(\mv{q},\mv{\bar{w}}_k+\mv{\delta}_k)/2}}.
%\end{align}
As a result, the total RF energy received by each ER $k\in\mathcal K$ during the whole charging duration is expressed as
\begin{align}
E_k(\{\mv{q}(t)\},\mv{w}_{k})=\int_0^T Q_k(\mv{q}(t),\mv{w}_{k}) dt.
\end{align}

Our objective is to maximize the worst-case minimal energy transferred to all ERs over the whole charging duration, by optimizing the UAV's positioning locations $\{\mv{q}(t)\}$ over time, subject to the ERs' bounded location errors.
Therefore, the problem of our interest is formulated as
\begin{align}
\text{(P1)}:\max_{\{\bm{q}(t)\}}~\min_{k\in \mathcal{K}}~\min_{\bm{w}_{k}\in\mathcal A_{k}} &~E_k(\{\mv{q}(t)\},\mv{w}_{k})\nonumber.
\end{align}
By introducing an auxiliary variable $E$, problem (P1) can be equivalently expressed as
\begin{align}
\text{(P1.1)}:\max_{\{\bm{q}(t)\},E}&~~~E\nonumber\\
\text{s.t.}&\min_{\bm{w}_{k}\in\mathcal A_{k}} ~~E_k(\{\mv{q}(t)\},\mv{w}_{k})\geq E,\forall k\in \mathcal{K}.
\end{align}
Notice that optimally solving problem (P1) or (P1.1) is very challenging in general. First, (P1) or (P1.1) is a non-convex optimization problem as the objective function in (P1) is not concave or the constraints in (4) are non-convex. Second, the uncertainty in ERs' locations brings an infinite number of constraints in (4) that are difficult to be dealt with. Third, the energy function $E_k(\{\mv{q}(t)\},\mv{w}_{k})$ is generally not a continuous function under our generic path loss model.
\vspace{-10pt}\section{Proposed Solution to Problem (P1) or (P1.1)}
In this section, we propose an efficient algorithm to solve (P1.1). We first deal with the uncertainty in ERs' locations $\mv{w}_{k}$'s, and then obtain the optimal solution of transformed problem by applying the Lagrange duality method.
%\subsection{Robust Optimization}

First, we deal with the uncertainty issue on location variable $\mv{w}_{k}$ of each ER $k$, so as to transform the left-hand-side of constraint (4) into deterministic functions.
%Problem (P1.1) is still hard to be solved due to the uncertainty variable $\mv{w}_{k}$.
%However, it is known that the distance between the UAV and the ER $k\in\mathcal{K}$ increases, the UAV will experience a higher degree of LoS obstruction with $\alpha(\mv{q},\mv{w}_k)$ increasing and $\beta(\mv{q},\mv{w}_k)$ decreasing.
This, however, is a difficult task, as each ER's location $\mv{w}_{k}$ is coupled with the UAV's trajectory $\{\mv{q}(t)\}$. To tackle this challenge, we first obtain the minimum of $Q_k(\mv{q}(t),\mv{w}_{k})$ under any given $\mv{q}(t)$ at time $t$, i.e.,
\begin{align}\label{A}
\min_{\bm{w}_{k}\in\mathcal A_{k}} Q_{k}(\mv{q}(t),\mv{w}_k).
\end{align}
 Notice that under any given ER location $\mv{w}_{k}$, the path loss exponent $\alpha(\mv{q}(t),\mv{w}_{k})$ and the reference path loss $\beta(\mv{q}(t),\mv{w}_{k})$ are monotonically increasing and decreasing with respect to the distance $d(\mv{q}(t),\mv{w}_k)$, respectively. Thus, under any given $\mv{q}(t)$, problem (5) is equivalent to maximizing $d(\mv{q}(t),\mv{w}_k)$, i.e.,
\begin{align}
\max_{\bm{w}_{k}\in\mathcal A_{k}}
\|\mv{q}(t)-\mv{w}_{k}\|.
\end{align}
 It is easy to verify that the optimality of problem (6) is attained when $\mv{w}_{k}^*(\mv{q}(t))=\mv{\bar{w}}_k+\epsilon\frac{\bm{\bar{w}}_k-\bm{q}(t)}{\|\bm{\bar{w}}_k-\bm{q}(t)\|}$, which means that the worst-case location $\mv{w}_{k}^*(\mv{q}(t))$ is $\epsilon$ meters far away from the approximated location $\mv{\bar{w}}_k$ in the direction of $\frac{\bm{\bar{w}}_k-\bm{q}(t)}{\|\bm{\bar{w}}_k-\bm{q}(t)\|}$. By substituting $\mv{w}_{k}^*(\mv{q}(t))$ into (5), the worst-case power transferred to each ER $k$ at time $t$ is thus expressed as
\begin{align}
\hat{Q}_k(\mv{q}(t))
&=\frac{\beta(\mv{q}(t),\mv{w}_{k}^*(\mv{q}(t)))P}
{(H^2+\|~\|\mv{q}(t)-\mv{\bar{w}}_k\|+\epsilon\|^2)^{\alpha(\bm{q}(t),\bm{w}_{k}^*(\bm{q}(t)))/2}}.
\end{align}
 By replacing $\min\limits_{\bm{w}_{k}\in\mathcal A_{k}}E_k(\{\mv{q}(t)\},\mv{w}_{k})$ in (4) as $\int_0^T\hat{Q}_k(\mv{q}(t))dt$, we thus transform problem (P1.1) as
\begin{align}
\text{(P2):}\max_{\{\bm{q}(t)\},E}~~~&E\nonumber\\
\text{s.t.}&\int_0^T\hat{Q}_k(\mv{q}(t))dt\geq E,\forall k\in \mathcal{K}.
\end{align}
%\subsection{Proposed Solution to (P2)}

Next, we proceed to solve problem (P2).Though problem (P2) is still non-convex with non-continuous and non-convex constraint functions, it satisfies the so-called time-sharing condition in \cite{yu2006dual}. Thus, the strong duality holds between (P2) and its Lagrange dual problem. As a result, we solve problem (P2) optimally according to the Lagrange duality method \cite{Boyd:2004:CO:993483}. Notice that a similar approach has been used for solving the minimum energy maximization problem under infinite charging duration and free-space channel model in\cite{xu2018uav}.

Let $\lambda_k\geq0$, $k\in \mathcal{K}$, denote the dual variable associated with the $k$-th constraint in (8). Then the Lagrangian of (P2) is
\begin{align}
\mathcal{L}(\mv{q}(t),E,\{\lambda_k\})=(1-\sum\limits_{k\in\mathcal{K}}\lambda_k)E+
\int_0^T\sum_{k\in\mathcal{K}}\lambda_k\hat{Q}_k(\mv{q}(t))dt.
\end{align}
Accordingly, the dual function of (P2) is
\begin{align}
g(\{\lambda_k\})=\max_{\{\bm{q}(t)\},E}\mathcal{L}(\mv{q}(t),E,\{\lambda_k\}).
\end{align}
As $\sum_{k\in\mathcal{K}}\lambda_k=1$ must hold for $g(\{\lambda_k\})$ to be bounded from above, the dual problem of (P2) is
\begin{align}
\text{(D2):}\min_{\{\lambda_k\ge 0\}}~&g(\{\lambda_k\})\nonumber\\
\mathrm{s.t.}&\sum_{k\in\mathcal{K}}\lambda_k=1.
\end{align}
  In the following, we first solve problem (10) under any given feasible  $\{\lambda_{k}\}$ to obtain $g(\{\lambda_k\})$, then  find the optimal $\{\lambda_{k}\}$ to minimize $g(\{\lambda_k\})$, and finally construct the optimal primal solution to (P2).

Under any given feasible $\{\lambda_{k}\}$, we first decompose problem (10) into the following two sets of subproblems.
\begin{align}
&\max_{E}~(1-\sum\limits_{k\in\mathcal{K}}\lambda_k)E\\
&\max_{\bm{q}(t)}~\sum_{k\in \mathcal{K}}\lambda_k\hat{Q}_k(\mv{q}(t)), \forall t\in (0,T]
\end{align}
 As for subproblem (12), since $\sum_{k\in\mathcal{K}}\lambda_k=1$ holds, its  objective value is always zero, and thus we can arbitrary choose any real number as the optimal (but non-unique) $E^{\{\lambda_k\}}$. On the other hand, notice that the subproblems in (13) are identical for all $t\in(0,T]$. In this case, we adopt a two-dimensional (2D) exhaustive search over the region $[\underline{x},\overline{x}]\times[\underline{y},\overline{y}]$, with $\underline{x}=\min\limits_{k\in\mathcal{K}}x_k$, $\overline{x}=\max\limits_{k\in\mathcal{K}}x_k$,
$\underline{y}=\min\limits_{k\in\mathcal{K}}y_k$ and $\overline{y}=\max\limits_{k\in\mathcal{K}}y_k$ to find the optimal solution as $\mv{q}^{\{\lambda_{k}\}}$.
By substituting $\mv{q}^{\{\lambda_{k}\}}$ into (10), the dual function $g(\{\lambda_k\})$ is obtained. Notice that if the optimal solution of $\mv{q}^{\{\lambda_k\}}$ to (13) is
 non-unique, we can arbitrarily choose any one of them to obtain the dual function.

 Next, with $g(\{\lambda_k\})$ obtained, we solve the dual problem (D2). As the dual function $g(\{\lambda_k\})$ is always convex but generally non-differentiable, we solve (D2) by utilizing subgradient-based methods, such as the ellipsoid method \cite{ellipsoid}. We denote the obtained optimal dual solution to (D2) as $\{\lambda_{k}^\star\}$ and corresponding optimal positioning as $\mv{q}^{\{\lambda_{k}^{\star}\}}$.
% The subgradient of the objective function $g(\{\lambda_k\})$ is given by $\mv{s}_0(\lambda_1,...,\lambda_K)$=$[T Q_1(\mv{q}^{\{\lambda_k\}}),...,T Q_K(\mv{q}^{\{\lambda_k\}})]$, and the equality constraint in (10) we can seen as two inequality constraints ($1-\sum_{k\in\mathcal{K}}\lambda_k\leq0$,$1-\sum_{k\in\mathcal{K}}\lambda_k\geq0$), whose subgradients are given by $\mv{s}_1(\lambda_1,...,\lambda_K)=-\mv{e},\mv{s}_2(\lambda_1,...,\lambda_K)=\mv{e}$, respectively, with $\mv{e}$ denoting an all-one vector.

  Based on the optimal $\{\lambda_{k}^\star\}$, we still need to construct the optimal solution to (P2). If the optimal solution $\mv{q}^{\{\lambda_{k}^{\star}\}}$ is unique, it is also the optimal solution to  problem (P2), and the corresponding optimal minimum-energy is $E^\star=\hat{Q}_k(\mv{q}^{\{\lambda_{k}^{\star}\}})T$. Otherwise, we need to time-share among these optimal solutions by allowing the UAV to hover above each location with a certain duration to be optimized. Suppose that there are $\Gamma\geq1$ optimal UAV positioning solutions to problem (13) under $\{\lambda^\star\}$, denoted by  $\{\mv{q}_\gamma^{\star}\}_{\gamma=1}^\Gamma$, and let $\tau_\gamma$ denote the hovering duration at $(\mv{q}_\gamma^\star,H)$. When the UAV stays at $\mv{q}_\gamma^\star$, let $\hat{Q}_k(\mv{q}_\gamma^\star)$ denote the received RF power at each ER $k\in \mathcal{K}$. Then, the optimal $E^\star$ and $\{\tau_\gamma^\star\}$ can be obtained by solving the following problem.
\begin{align}
\text{(P3):}\max_{\{\tau_{\gamma}\ge 0\},E}~~&E\nonumber\\
\mathrm{s.t.}&\sum_{\gamma=1}^{\Gamma}\tau_\gamma \hat{Q}_k(\mv{q}_\gamma^\star)\geq E,\forall k\in\mathcal{K}\\
&\sum_{\gamma=1}^{\Gamma}\tau_\gamma=T.
\end{align}
Problem (P3) is a linear program, which can thus be solved by using the interior point method \cite{Boyd:2004:CO:993483}.

Finally, with $\{\tau_\gamma^\star\}$ and $E^\star$ at hand, we divide the whole charging duration $T$ into $\Gamma$ portions, denoted by periods $\mathcal{T}_1,...,\mathcal{T}_\Gamma$, where $\mathcal{T}_\gamma=(\sum_{i=1}^{\gamma-1}\tau_\gamma^\star,\sum_{i=1}^{\gamma}\tau_\gamma^\star]$ with duration $\tau_\gamma^\star, \gamma \in \{1,...,\Gamma\}$. Then, we have the obtained solution in the following proposition.
\begin{proposition}
The optimal solution to (P2) or the obtained solution to (P1) is given as
$\mv{q}^\star(t)=\mv{q}_{\gamma}^\star,~\forall t \in \mathcal{T}_\gamma,~\gamma\in\{1,...,\mathcal{T}\}$, and the correspondingly achieved minimum-energy is $E^\star$. It is evident that the obtained trajectory has an interesting multi-location-positioning structure, i.e, the UAV should be positioned among different locations over time, instead of always staying at a single location.
\end{proposition}
\begin{figure}
  \centering
  \includegraphics[width=5.5cm]{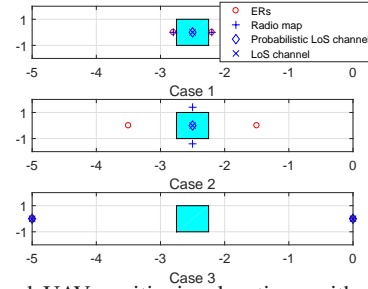}
  \vspace{-1.5em}
    \small\caption{\small  Optimized UAV positioning locations with $K=2$ ERs, in which the blue cube denotes an obstacle.}
    \label{fig1}
     \vspace{-2em}
\end{figure}
\vspace{-10pt}\section{Numerical Results}
This section provides numerical results to validate the performance of our proposed radio-map-based robust positioning design. In the simulation, we consider the segment propagation channel model \cite{J.Chen2017}, where if there are obstacles between the UAV and ER $k$, the channel follows a NLoS model with the path loss exponent being $\alpha_{\text{NLoS}}=4$ and reference channel power gain being $\beta_{\text{NLoS}}=10^{-4}$, and otherwise, the channel is  LoS with the path loss exponent being $\alpha_{\text{LoS}}=2.3$ and reference channel power gain being $\beta_{\text{LoS}}=10^{-3}$. Unless otherwise stated, we set the UAV altitude as $H=\text{5 m}$, the maximum transmit power as $P=\text{10 W}$ and the prespecified ER location error as $\epsilon=\text{1 m}$. For performance comparison, we also consider the conventional LoS channel and probabilistic LoS channel, where we set the channel parameters as $\alpha_0 = 2$, $\beta_0=10^{-3}$, $A=10$, $B=0.6$ and $\eta=0.1$, which approximately match our obstacles' distribution (please refer to \cite{al2014optimal} on the details of the channel parameters).
%\subsection{Optimal hovering positioning in two ERs case}

\begin{figure}
  \centering
  \includegraphics[width=5.5cm]{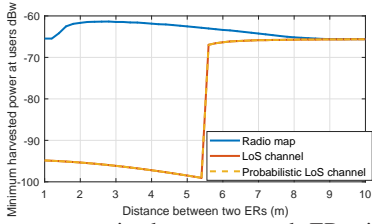}
  \vspace{-1em}
    \small\caption{\small The common received power at each ER with $K=2$ ERs.}
    \label{fig2}
     \vspace{-2em}
\end{figure}
First, we consider the case with $K=2$ ERs as shown in Fig. \ref{fig1}, where there is one obstacle (i.e., the blue cube) between the two ERs with height 4.5 m. It is observed that when the two ERs' distance is short (e.g., Cases 1 and 2), the optimized UAV positioning locations under the radio-map based design are different from those based on conventional LoS and probabilistic LoS models, in order to avoid the link blocking due to obstacles. By contrast, when the ERs' distance becomes large (e.g., Case 3), the three approaches are observed to lead to similar positioning locations.
%In Cases 1 and 2, it is also observed that there are no obstacles blocking the links between the UAV and ERs under the radio map channel, while in LoS and probabilistic LoS channel the links between the UAV and ERs are blocked by the obstacle, which comprises ERs' received power.
 Fig. \ref{fig2} shows the average minimum or common received power versus the ERs' distance. It is observed that when the distance is sufficiently large, the performances under the three schemes are almost same. When the distance becomes shorter, our proposed radio-map-based channel is observed to considerably outperform the conventional designs with LoS and probabilistic LoS channels. This is consistent with Fig. 1, in which our proposed design can efficiently find UAV positioning locations with LoS links to ERs, by exploiting the exact channel propagation information based on the radio map.
%Moreover, it can also be observed that radio-map-based channel show its superior advantage in realistic scene while some obstacles leading serious block to LoS channel.
%\subsection{Optimal hovering positioning in Multi ERs case}

\begin{figure}
  \centering
  \includegraphics[width=5.5cm]{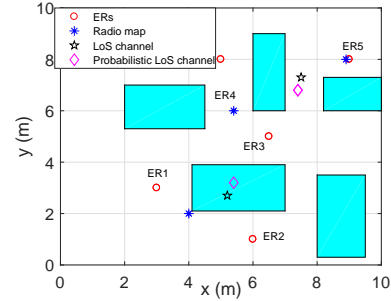}
  \vspace{-1em}
    \small\caption{\small Optimized UAV positioning locations with $K=5$ ERs, in which blue cubes denote obstacles. }
    \label{fig3}
\end{figure}

\begin{figure}
  \centering
  \includegraphics[width=5.5cm]{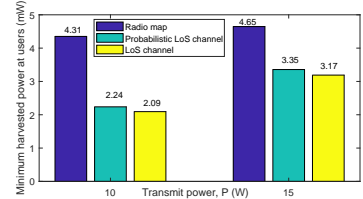}
  \vspace{-1em}
    \small\caption{\small The max-min average received power at each ER versus the transmit power P with $K=5$ ERs.}
    \label{fig4}
     \vspace{-2em}
\end{figure}

Next, we consider the setup with $K=5$ ERs located in an area of $10\times10$ square meters as shown in Fig. \ref{fig3}, in which the blue cubes denote the obstacles (e.g., trees) with height $\text{4.5 m}$.
% As for radio-map-based channel, we consider only two segments (i.e. LoS and NLoS segments) for each ER $k\in \mathcal{K}$ in the simulation. Corresponding channel parameters are chosen as Table \uppercase\expandafter{\romannumeral1}.
It is observed that under the radio-map-based design, there are three optimal positioning locations, which are close to ERs 1-2, ERs 3-4, and ER 5 with LoS connections, respectively. By contrast, under the LoS/probabilistic LoS channel models, the links between the UAV and ground ERs 1-4 are observed to be blocked by obstacles, resulting in severe performance loss. This is because either the deterministic LoS or the stochastic probabilistic LoS channel models cannot capture the specific features of the real channel environment, which comprises the performance.
%On the other hand, Fig. 3 shows that even under probabilistic LoS channel model, the optimal hovering positioning are blocked by obstacles, as the model is a stochastic one that cannot capture the exact environment.
%On the one hand, it is easily observed that UAV will hover above a middle point in which some ERs (e.g.,ERs1-2,ERs3-4) are close to each other while holding a LoS link to the UAV simultaneously.
%On the other hand, UAV will hover above a point close to the ER (e.g.,ER5) when the ER is surrounded by some obstacles without any LoS link to the other ERs.
 Fig. \ref{fig4} shows the average minimum energy transferred to the $K$ ERs versus different transmit power $P$. It is observed that radio-map-based approach achieves approximately $100\%$ performance gain over the other two approaches in the case with $P=\text{10 W}$ and $40\%$ in the case with $P =\text{15 W}$.

 %Furthermore, the optimal hovering points of probabilistic LoS channel are closer to the middle point of all ERs to obtain a larger elevation angle compared to LoS channel.
\vspace{-15pt}\section{Conclusion}
This letter studied a new radio-map-based robust positioning optimization approach for a UAV-enabled multiuser WPT system.
 We maximized the minimum energy transferred to all ERs over a sufficiently long charging period, by assuming that the UAV only partially knows the ER's locations.
We proposed to use the radio map technique to acquire the channel propagation information, and then adopted the robust optimization and Lagrange duality method to solve this problem efficiently. Numerical results showed that our proposed radio-map-based design achieved significant performance gains against other benchmark designs under LoS and probabilistic LoS channels. This is due to the fact that our proposed design can better exploit the channel propagation environment information, while the LoS/probabilistic LoS models generally mismatch with the real radio environment.

%\bibliographystyle{IEEEtran}      %LaTex Class文件, IEEEtran 为给定模板格式定义文件名
%\bibliography{Ref}

\end{document}